\documentclass[showpacs,12pt,preprintnumbers,amsmath]{revtex4}
\usepackage{graphicx} 
\usepackage{bm}

\begin{document}

\title{Liquid-liquid transition in supercooled silicon determined by
first-principles simulation}

\author{P. Ganesh}
\affiliation{Carnegie Institution of Washington, Washington DC}
\author{M. Widom}
\affiliation{Department of Physics, Carnegie Mellon University, Pittsburgh, PA}

\date{\today}

\begin{abstract}
First principles molecular dynamics simulations reveal a liquid-liquid
phase transition in supercooled elemental silicon.  Two phases coexist
below $T_c\approx 1232K$.  The low density phase is nearly
tetra-coordinated, with a pseudogap at the Fermi surface, while the
high density phase is more highly coordinated and metallic in nature.
The transition is observed through the formation of van der Waals
loops in pressure-volume isotherms below $T_c$.
\end{abstract}

\maketitle

Silicon occupies a position in the periodic table at the border
between metals and insulators.  At low pressure, crystalline silicon
(c-Si) is tetrahedrally coordinated (like diamond) and is an indirect
band-gap semiconductor.  As pressure increases, crystalline Si
transforms from the diamond phase through a more highly-coordinated
metallic $\beta$-tin phase to a hexagonal phase.  Like c-Si,
noncrystalline amorphous silicon (a-Si) is semiconducting at low
pressure and maintains a coordination number $N_c\approx 4$.  At
higher pressures, amorphous silicon becomes metallic and more highly
coordinated.  Thus, both c- and a-Si exhibit pressure-induced
polymorphism.

Liquid silicon (l-Si) is metallic at high temperature, although it
possesses a small population of covalent
bonds~\cite{SiParrinello,SiHafner1}.  The coordination number
$N_c\approx 7.3$ (for liquid and amorphous structures $N_c$ is the
average number of neighbors within the first peak of the radial
distribution function $g(r)$) is lower than occurs in ordinary metals
such as aluminum or copper, but higher than in c-Si.  As expected from
its higher coordination number, l-Si has higher density than c-Si, a
property that silicon shares with water~\cite{Water}.

The structure of supercooled l-Si and its transition from a metallic
dense-packed structure at high temperature to a semiconducting open
network at low temperature remains uncertain.  The dissimilarity of
the amorphous and liquid states raises the possibility of intermediate
phases~\cite{Aptekar}.  By analogy with the liquid-liquid phase
transition (LLPT) in water~\cite{Poole1992,Mishima1998}, where
tetrahedral order is also present in the crystalline solid phase,
researchers speculate~\cite{Poole1997} such a transition might occur
in silicon, between a tetracoordinated low density liquid (LDL) and a
more highly coordinated high density liquid (HDL).  Since the LLPT
presumably occurs in the supercooled liquid, both LDL and HDL are
metastable states, if they exist at all.  The supercooled HDL state is
the metastable extension of the high temperature equilibrium liquid
(HTEL) state, while the glassy a-Si state is the extension of LDL to
low temperature.

Experimental evidence for an LLPT is not conclusive.  Pressure induced
transformation of a-Si~\cite{SiZulu}, from a low-density amorphous
(LDA) state at low pressure to a high-density amorphous state (HDA),
similar to that of amorphous ice~\cite{PressIce}, suggests the
possibility of an LLPT in silicon similar to that in supercooled
water.  Plastic deformation~\cite{Hedler2004} studies indicate a
transition from a-Si to a fluid state, tentatively identified as LDL,
at a temperature around 1000K (the melting point is $T_m=1687K$).
Containerless supercooling
experiments~\cite{CNL1,CNL2,EML1,EML2,Goldman2005} lead to
contradictory results regarding the temperature dependence of the
supercooled liquid properties.

Computer simulations reveal complete structural detail not available
by ordinary experimental means.  However, they suffer from uncertainty
arising from approximate models of interatomic forces and energetics,
as well as limitations of sample size and simulation time.  To capture
the nature of tetrahedral bonding, the Stillinger-Weber (S-W)
potential~\cite{SWpot} includes angle dependent interactions.  This
potential shows clear evidence of a first-order HDL to LDL transition,
in the deeply supercooled regime (with a transition temperature $T_c
\sim 1060K$)~\cite{Sastry}.  Electronic structure calculations on
these simulated configurations report a metal to semimetal transition
across the LLPT transition temperature~\cite{SemitoMetal}.  Other
empirical potentials, such as the Keating potential~\cite{Keating} and
the environment dependent potential~\cite{EDIP}, find contradictory
results, so the nature and existence of an LLPT depends on the form of
the interaction model~\cite{llclasspot}.

\begin{figure}
\includegraphics*[width=5in,angle=-90]{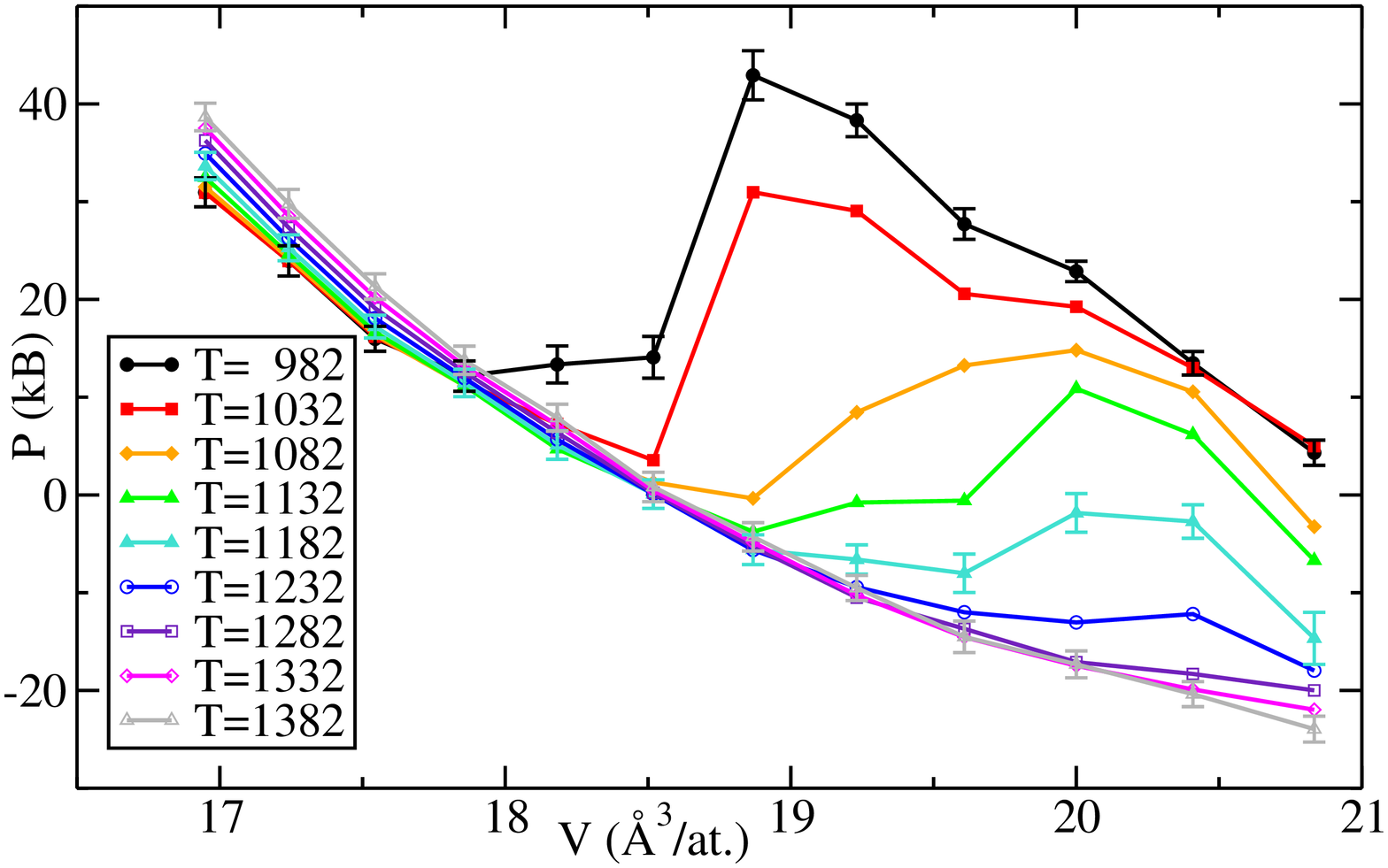}
\caption{\label{fig:PVT} (color online) Pressure-volume isotherms of
liquid Si.  Data points are calculated using Eq.~\ref{eq:PofT}.}
\end{figure}

Since neither experiments nor simulations with empirical potentials
resolve the existence or character of an LLPT in Si with certainty,
further investigation using first-principles methods is warranted.
The pioneering Car-Parrinello simulations~\cite{SiParrinello,Car1985} studied
solid c-Si and a-Si in addition to high temperature l-Si.  A later
simulation of the supercooled liquid~\cite{Morishita} in a small
$N=64$ atom cell found increased tetrahedral order and a sudden drop
in $N_c$ at T=1100K.  Our study carries out a more extensive
investigation and finds van der Waals loops in the pressure-volume
isotherms (see Fig.~\ref{fig:PVT}), a strong indication of an LLPT,
with a critical point around $T_c\approx$1232K.

Our simulations use the Vienna Ab-Initio Simulation Package
(VASP)~\cite{VASP,VASP2} together with Projector Augmented Wave (PAW)
potentials~\cite{PAW,KJ_PAW} in the Generalized Gradient Approximation
(GGA).  Electronic structure calculations of energies, pressures and
forces used the $\Gamma$ $k$-point only and were conducted at the
default energy cutoff of 245eV.  Molecular dynamics utilized a 2fs
time step, and velocities were rescaled every time step to maintain
constant temperature.  Our molecular dynamics calculations extended to
3ps or more and required 2 weeks of computer time for each temperature
and volume.

\begin{figure}
\includegraphics*[width=5in,angle=-90]{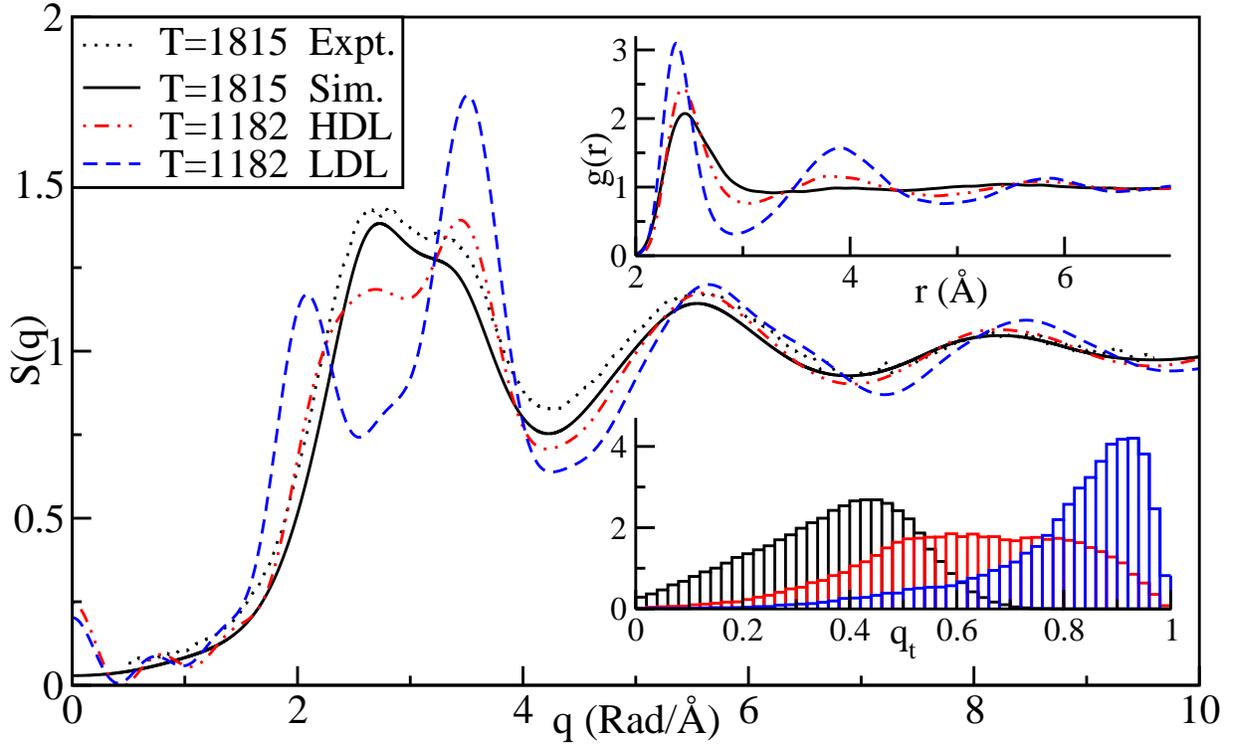}
\caption{\label{fig:HT-LT} (color online) Liquid structure factors
$S(q)$, radial distribution functions $g(r)$ and tetrahedral order
parameter ($q_t$) distributions.  High temperature simulation at
temperature $T$=1815K and volume per atom $v=18.2$~\AA$^3$ is compared
with experiment~\cite{Kelton}.  Low temperature simulation at
$T$=1182K shows coexisting structures, HDL at $v$=18.9~\AA$^3$ and LDL
at $v$=20.4~\AA$^3$.}
\end{figure}

To test for finite size effects we evaluated the radial distribution
function $g(r)$ at system sizes $N$=100, 200 and 300 and found
adequate convergence at $N=200$ atoms.  We transformed $g(r)$ to
obtain the structure factor $S(q)$, using the Baxter
method~\cite{Ganesh08} (see Fig.~\ref{fig:HT-LT}).  The peak
and shoulder positions agree well with experiment~\cite{Kelton}, but
the experimental $S(q)$ exhibits a slight excess at low $q$ arising
from a background effect, causing the experimental data to violate a
sum rule associated with short distance correlations~\cite{Ganesh08},
while the simulation data obey the sum rule.

Ideally a phase transition should be observed as a function of
temperature $T$ and pressure $P$.  However, the VASP program is
restricted to fixed volume simulation, so we monitored the reported
``external pressure'' (derivative of potential energy with respect to
volume~\cite{Martin1}) and added the kinetic energy contribution $\rho
k_B T$ to obtain the thermodynamic pressure.  We also add a constant
offset of $P_{Pulay}$=1.4 kilobar to account for the Pulay stress,
which we found to be reasonably independent of temperature and volume.

At least 9 temperatures were studied at each volume, uniformly spaced
between a low of T=982K and a high of 1382K, with additional
intermediate temperatures included in cases where the mean internal
energy varied rapidly with temperature.  Each starting configuration
for a given V and T was generated by a lengthy MD simulation prior to
the beginning of data collection.  To improve the efficiency of
configuration sampling, we employed a replica exchange
method~\cite{REM,NewmanBarkema,Ganesh08} to swap configurations
between temperatures.  Briefly, pairs of configurations at common
volume but differing in energy by $\Delta E$ and differing in inverse
temperature ($\beta=1/k_BT$) by $\Delta\beta$, were swapped with
probability $\exp{(\Delta\beta\Delta E)}$.  We attempted swaps every
100fs.
  
Having energy and pressure data at a series of nearby temperatures
allows us to employ the ``multiple histogram''
technique~\cite{MultiHisto,NewmanBarkema}.  Histograms of internal
energy at fixed temperature $H_T(E)$ are converted into
configurational densities of states $\Omega(E)=(\sum_T H_T(E))/(\sum_T
e^{(F(T)-E)/k_BT})$ where the sums run over the discrete temperatures
$T$ at which simulations were performed.  Consistency between
temperatures is enforced by their free energies $F(T)=-k_BT\ln{Z(T)}$
where the partition function $Z(T)=\int\Omega(E)e^{-E/k_BT}{\rm d}E$.
We thus reconstruct the free energy and its derivative quantities such
as pressure and heat capacity as analytic functions of temperature.
For example the pressure is obtained from
\begin{equation}
\label{eq:PofT}
P(T)=\frac{1}{Z(T)}\int\bar{P}(E)\Omega(E)e^{-E/k_BT}{\rm d}E
\end{equation}
where $\bar{P}(E)$ is the mean pressure observed at energy $E$.
Accuracy of this method depends on a thorough sampling of
configuration space and on sufficient overlap of the energy histograms
at adjacent temperatures.

Results for the pressure are illustrated in Fig.~\ref{fig:PVT}.  Error
bars are estimates of standard error, defined as the RMS fluctuation
in pressure divided by $\sqrt{N_{ind}}$, where we set $N_{ind}$=10 as
the approximate number of independent samples.  Two main features of
this plot are: van der Waals loops, and negative thermal expansion.

Van der Waals loops occur when a region of positive slope interrupts
the generally negative slope of the pressure-volume isotherm.
Positive slope of $P(V)$ corresponds to negative isothermal
compressibility ($K_T=-\frac{1}{V}(\frac{\partial V}{\partial P})_T$).
This thermodynamically unstable state would phase separate into high
volume (low density) and low volume (high density) phases if the
system size were sufficiently large.  However, negative $K_T$ is
permissible in systems of finite size owing to the free energy cost of
the interface needed to separate the two phases.  Maxwell's equal area
construction can be used to identify the coexisting states below the
critical temperature, which we identify as $T_c\approx 1232$K.  At
$T=1182$K, we find coexisting states: HDL at density
$\rho=0.053$~atoms/\AA$^3$ (volume $v=18.9$~\AA$^3$/atom) with
coordination number $N_c=5.6$; LDL at density $\rho=0.049$ ($v=20.4$)
with coordination number $N_c=4.2$.  According to the isotherms, HDL
is the extension of the equilibrium liquid.

Negative thermal expansion ($\alpha_P=\frac{1}{V}(\frac{\partial
V}{\partial T})_P$) occurs when the pressure is a {\em decreasing}
function of temperature and can be seen at volumes above
18~\AA$^3$/atom.  Negative thermal expansion is thermodynamically
permissible, and is seen in water close to freezing, for example, but
is quite rare.  Thermal expansion is related to entropy via $(\frac{\partial
S}{\partial V})_T=\frac{\alpha_P}{K_T}$.  Since LDL exhibits negative
$\alpha_P$, we see that entropy {\em decreases} (counterintuitively)
as volume increases.  Presumably this reflects the higher degree of
orientational order in the open tetracoordinated network as compared
with the disordered high density liquid state~\cite{Debenedetti03}.

Fig.~\ref{fig:HT-LT} (inset) shows the distribution of the tetrahedral order
parameter~\cite{Water}, 
\begin{equation}
q_t=1-\frac{3}{8}\sum_{i<j=1}^4(\cos{\theta_{ij}}+\frac{1}{3})^2
\end{equation}
where $\theta_{ij}$ is the angle formed by an atom with its $i^{th}$
and $j^{th}$ nearest neighbors.  A value close to 1 indicates
tetrahedral bonding.  We evaluate this distribution in the HDL and LDL
states at temperature T=1182K.  At lower densities, the favored LDL
structure is indeed much more tetrahedral than the higher density HDL
or HTEL structures.

Because the LLPT occurs between metastable liquid phases, below the
equilibrium freezing transition, a crucial question is if our
coexisting phases at low temperature are both truly liquid.  Indeed,
since LDL exhibits fairly strong peaks in $g(r)$ and $S(q)$ as well as
a high degree of tetrahedral order, a central concern is if this LDL
phase has not actually crystallized.  Likely candidates for the
crystal structure are the usual c-Si diamond structure of Pearson type
cF8, or the Pearson type cI16 crystal that arises upon recovery at
atmospheric pressure following high pressure treatment.

\begin{figure}
\includegraphics*[width=5in,angle=-90]{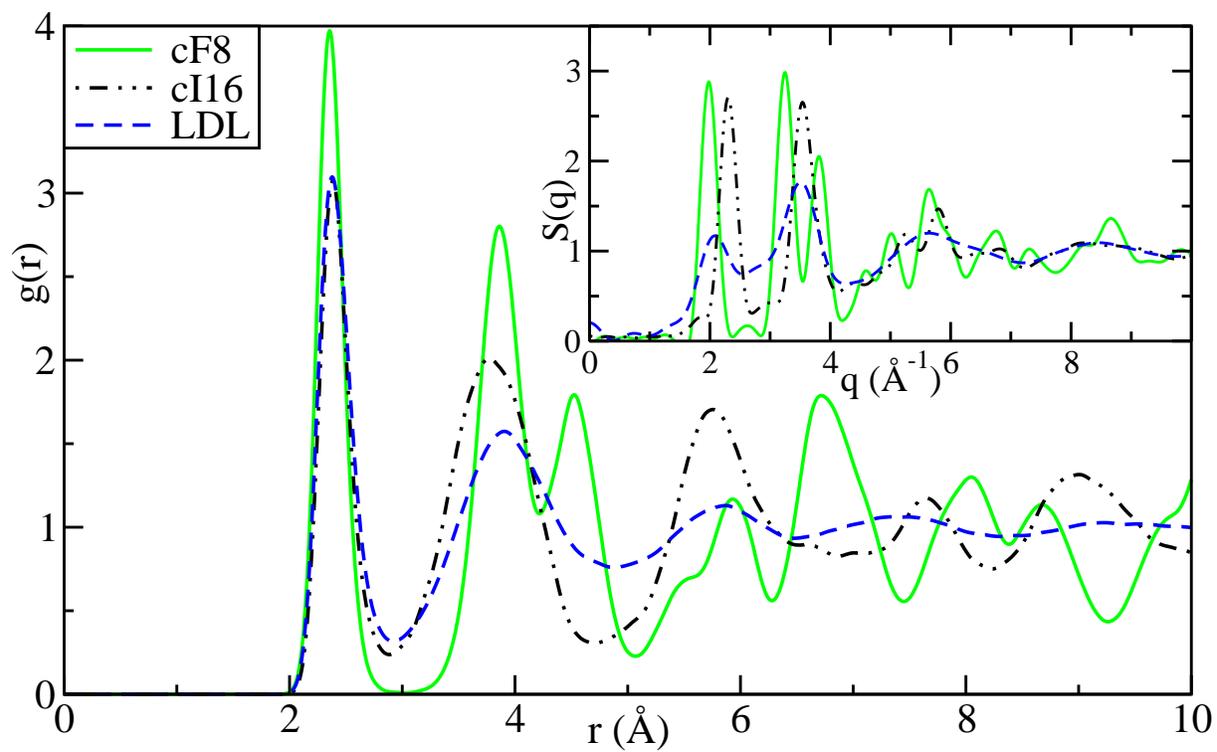}
\caption{\label{fig:LDL-vs-xtal} (color online) Comparison of
low-density liquid structure with competing crystal structures at
$T=1182$K.}
\end{figure}

To make direct comparisons we carried out solid-state molecular
dynamics simulations.  For cF8 we took $N$=216 atoms at
$v$=20.4~\AA$^3$, and for cI16 we took $N$=192 atoms at
$v$=18.5~\AA$^3$, the densities being chosen taking thermal expansion
into account.  Comparisons of $g(r)$ and $S(q)$ are shown in
Fig.~\ref{fig:LDL-vs-xtal}.  In order to smear out the Bragg peaks in
$S(q)$ we smoothly truncated the crystalline $g(r)$ prior to Fourier
transformation, so the figures understate the strength of peaks in
$S(q)$ for the crystalline structures.  At $T$=1182K both crystal
structures remain far better ordered than the LDL.  Additionally, for
both $g(r)$ and $S(q)$ there is no systematic resemblance between LDL
and either crystal structure.

Comparing $q_t$ distributions at T=1182K (not shown) we found the
distribution for cF8 was more sharply peaked close to $q_t$=1 than that for
LDL, but the distribution for cI16 was rather similar to that of LDL.
Coordination numbers also were similar between LDL and cI16.  We
examined the statistics of atom rings.  For the ideal crystal
structures, both cF8 and cI16 have only even length rings, and no
rings of length less than 6.  For purposes of comparison we take the
ratio of the number of rings with length 5 or 7 to the number with
length 6 or 8.  Values of $(N_5+N_7)/(N_6+N_8)$ at $T$=1182K were:
0.00 for cF8, 0.08 for cI16 and 0.50 for LDL.  We also found some
rings of length 3 and 4 in LDL, but not in either cF8 or cI16.
Finally, we visually inspected atomic configurations and saw no hints
of crystallinity in LDL

\begin{figure}
\includegraphics*[width=5in,angle=-90]{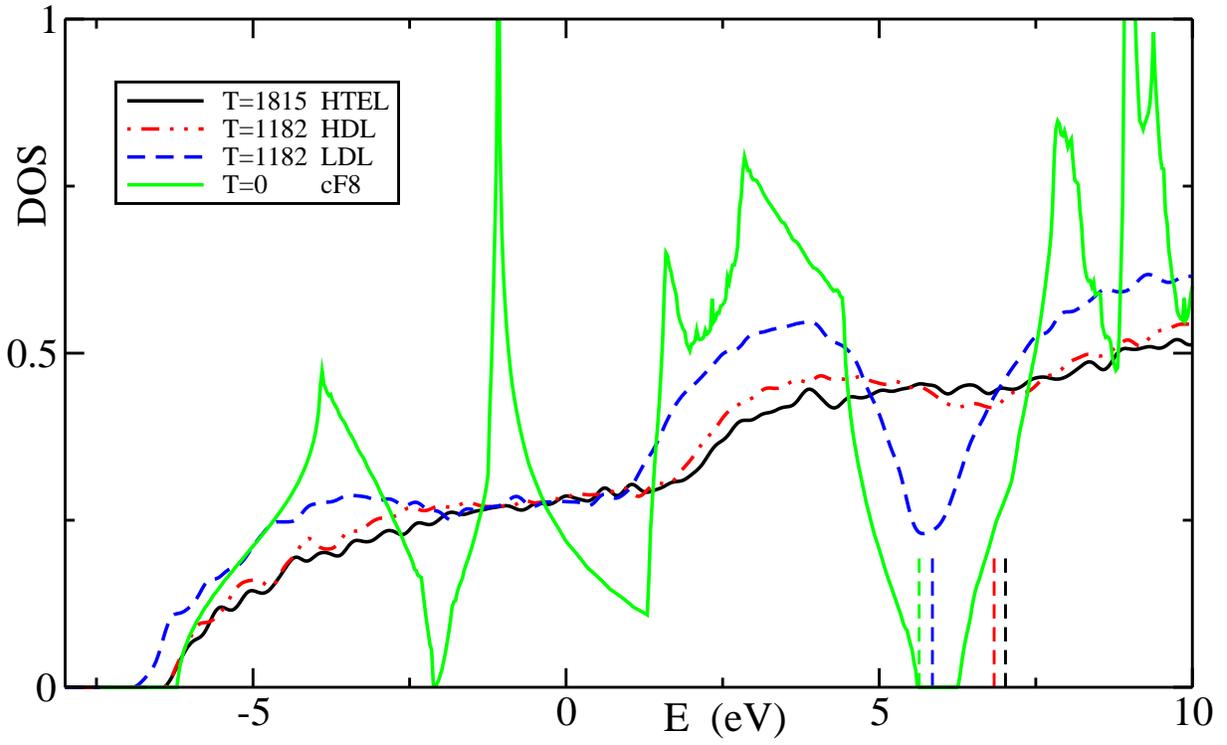}
\caption{\label{fig:DOS} (color online) Electronic densities of states
(states/eV/atom).  Vertical dashed lines locate $E_F$.}
\end{figure}

The electronic density of states (DOS) governs important material
properties including electric conductivity.  Fig.~\ref{fig:DOS}
illustrates the DOS of the equilibrium HTEL, metastable LDL and HDL,
and c-Si states.  Liquid state DOS are presented as the average of
five independent configurations.  The Fermi energies $E_F$ of HTEL and
HDL are similar because their atomic volumes are similar.  Likewise
the $E_F$'s are similar for LDL and cF8, but both lie below those of
HTEL and HDL because their atomic volumes are greater than for HDL and
HTEL.  The bottom of the LDL valence band (around -7eV) lies slightly
below cF8, reflecting the slightly higher coordination of LDL.  HTEL
and HDL have high densities of states at $E_F$ implying metallic
character.  LDL has a deep pseudogap at $E_F$ suggesting semimetallic
character~\cite{SemitoMetal}.  c-Si exhibits the characteristic
bandgap of a semiconductor.

In conclusion, we find that deeply supercooled liquid silicon
undergoes a liquid-liquid phase transition, separating into a high
density highly coordinated metallic liquid and a low density low
coordinated semimetallic liquid.  Our calculation reveals structural
detail not currently available from experiment in this temperature
range.  Because we employ first-principles forces, we have confidence
in the validity of the model.  The critical point lies close to
$T=1232$K, not far from estimates from
experiment~\cite{SiZulu,Hedler2004} and simulations based on empirical
potentials~\cite{Sastry}.  HTEL and HDL share characteristics such as
low atomic volume, high coordination, moderate $q_t$ and metallic
conduction, consistent with the notion that HDL is the metastable
extension of HTEL into the supercooled regime.  Likewise c-Si and LDL
share characteristics such as high atomic volume, low coordination,
large $q_t$ and low DOS at $E_F$, consistent with LDL exhibiting local
tetrahedral order similar to c-Si.

\begin{acknowledgments}
We wish to thank Bob Swendsen for numerous helpful discussions.  We
also thank Ken Kelton and Alan Goldman for sharing the experimental
data plotted in Fig.~\ref{fig:HT-LT}.
\end{acknowledgments}

\bibliography{si}

\end{document}